\documentclass[twocolumn,showpacs,prb]{revtex4}

\usepackage{graphicx}
\usepackage{dcolumn}
\usepackage{amsmath}
\begin{document}

\preprint{}


\title[Short Title]{Ultrahigh-Field Hole Cyclotron Resonance Absorption in InMnAs Films}



\author{Y. H. Matsuda}
\email{ymatsuda@cc.okayama-u.ac.jp}

\affiliation{Department of Physics, Faculty of Science, Okayama University, 3-1-1 Tsushimanaka, Okayama 700-8530, Japan}


\author{G. A. Khodaparast}
\author{M. A. Zudov}
\thanks{Present address: Physics Department, University of Utah,
Salt Lake City, Utah 84112.}
\author{J. Kono}
\email{kono@rice.edu}

\affiliation{Department of Electrical and Computer Engineering, Rice University, Huston, Texas 77005}

\author{Y. Sun}
\author{F. V. Kyrychenko}
\author{G. D. Sanders}
\author{C. J. Stanton}

\affiliation{Department of Physics, University of Florida, Gainesville, Florida 32611-8440}

\author{N. Miura}\author{S. Ikeda}\author{Y. Hashimoto}\author{S. Katsumoto}

\affiliation{Institute for Solid State Physics, University of Tokyo, Kashiwa, Chiba 277-8581, Japan}

\author{H. Munekata}

\affiliation{Image Science and Engineering Laboratory, Tokyo Institute of Technology,
Yokohama, Kanagawa 226-8503}

\date{\today}%


\begin{abstract}
We have carried out an ultrahigh-field cyclotron resonance study of $p$-type
In$_{1-x}$Mn$_x$As films, with Mn composition $x$ ranging from 0\% to 2.5\%,
grown on GaAs by low-temperature molecular-beam epitaxy.
Pulsed magnetic fields up to 500 T were used to make cyclotron resonance
observable in these low-mobility samples.
The clear observation of hole cyclotron resonance is direct evidence of the existence of
a large number of itinerant, effective-mass-type holes rather than
localized $d$-like holes.  It further suggests that the $p$-$d$ exchange mechanism
is more favorable than the double
exchange mechanism in this narrow gap InAs-based dilute magnetic semiconductor.
In addition to the
fundamental heavy-hole and light-hole cyclotron resonance absorption appearing
near the high-magnetic-field quantum limit, we observed
many inter-Landau-level absorption bands whose transition probabilities are strongly
dependent on the sense of circular polarization of the incident light.

\end{abstract}


\maketitle



\section{Introduction}
\label{sec:level1}
The successful growth of Mn-based III-V dilute magnetic semiconductors (DMSs) by
low temperature molecular beam epitaxy
\cite{munekata1,munekata0,ohnoPRL92,munekata2,HOhno_science} has opened up new
device possibilities to make simultaneous use of magnetic and semiconducting
properties.
Although there exist different types of theoretical models to explain the
microscopic mechanism of ferromagnetism in these semiconductors,
\cite{Matsukura,Dietl,Konig,Chattopadhyay,Akai,Inoue,Litvinov,Berciu,
Takahashi,Shirai,HKYoshida,Zhao,Jain}
a comprehensive picture has not been completed yet. Only an accepted fact 
at the present is that the ferromagnetic interaction between Mn ions is mediated by holes.

There are two limiting cases for the possible interaction mechanisms: the
Ruderman-Kittel-Kasuya-Yosida (RKKY) type interaction \cite{Matsukura}
and the double exchange (DE) type interaction.\cite{Akai} However,
neither of them is likely to provide a complete explanation.  The RKKY
interaction is inconsistent with the fact that the estimated magnitude
of the exchange interaction constant exceeds the Fermi energy as well as
the fact that ferromagnetism occurs even in the insulating phase.  On the
other hand, the DE mechanism has the problem that the valence states of most
of the Mn ions are found to be $d^5$ ($S=5/2$); there has been no strong
evidence that the $d$-holes which contribute to the DE interaction exist.
Therefore, the real interaction mechanism may be described by another model 
in which the character of the carrier is not pure $p$- or $d$-hole.

The Zener model, which is the limiting case of the RKKY model for small
carrier concentrations, is one of the possible mechanisms.\cite{Dietl}
In this case, if the $d$ state is deep enough in the valence band, holes at the Fermi level
have almost $p$-type character since the anti-bonding state formed by the
$p$-$d$ hybridization is almost $p$-like and we can classify the system
as a charge transfer insulator.\cite{Dietl}  However, there can be a
case where the $d$-state is shallow and the holes have partly $d$-like
character due to the strong $p$-$d$ hybridization. When the itinerant
holes are $d$-like, the reality is closer to the DE model.\cite{HKYoshida}
Therefore, investigating the degree of localization of holes is very
important in clarifying the physical picture of the carrier-induced
ferromagnetism.

Since cyclotron resonance (CR) directly provides the effective masses
and scattering times of carriers, we can obtain detailed information
on the itinerancy of the carriers.  In DC transport measurements it is
normally difficult to deduce the contributions of the mass and scattering
time independently. Hence, CR on III-V ferromagnetic semiconductors is
one of the most intriguing and urgently required experiments.  However,
since the mobilities of III-V ferromagnetic semiconductors are low ($<$ 100
cm$^2$/Vs), it is extremely difficult or impossible to observe CR in the
magnetic field range of a conventional superconducting magnet ($<$ 20 T).

Recently, we reporetd results of an ultrahigh field ($\leq$
100 T) CR study of {\it electrons} in In$_{1-x}$Mn$_x$As
films.\cite{nIMA-CR,Sanders2}  We described the dependence of cyclotron
mass on $x$, ranging from 0 to 12\%. We observed that the electron CR
peak shifts to lower field with increasing $x$. Midinfrared interband
absorption spectroscopy revealed no significant $x$ dependence of the band
gap. A detailed comparison of experimental results with calculations based
on a modified Pidgeon-Brown model allowed us to estimate the $s$-$d$
and $p$-$d$ exchange constants $\alpha$ and $\beta$ to be 0.5 eV and
$-$1.0 eV, respectively.

In this paper we report the first observation of {\em hole} CR in
$p$-type In$_{1-x}$Mn$_x$As films
using even higher magnetic fields ($\leq$ 500 T). The observation of CR
suggests that there are a large number of itinerant
holes in InMnAs and they behave like effective mass carriers rather than $d$-like
holes or holes tightly bound to the acceptor atoms.
This suggests that Zener's $p$-$d$ exchange interaction \cite{Dietl} is
more favorable as the ferromagnetic interaction mechanism than the DE interaction
in InAs-based magnetic semiconductors.

The observation of CR is also consistent with the Drude-like behavior observed in
the optical
conductivity of InMnAs thin films reported recently.\cite{Hirakawa1}
We also found that the observed CR
mass is not very different from that of holes in InAs.  We deduced the CR mobilities of
several samples with
different Mn concentrations from CR spectra at different photon
energies and compared them with
the values obtained by DC transport measurements.

Finally, we performed detailed theoretical calculations within the effective mass
formalism
using a modified Pidgeon and Brown model.\cite{PB,Sanders2} We found
that theoretical CR absorption spectra obtained by the simulations reproduce very well
the complicated CR spectra at very high fields.

%
%
%
\section{Experimental Procedure}
\label{sec:level2} Since the mobility of a ferromagnetic III-V
semiconductor is generally low (less than 100 cm$^2$/Vs), using
ultrahigh magnetic fields exceeding 100 T (megagauss field) is
essential for the present
study in order to satisfy the CR condition $\omega _c \tau  > 1$, where
$\omega_c$ is the cyclotron frequency and $\tau$ is the scattering
time.\cite{CMT-CR, nIMA-CR} Two kinds of pulsed
magnets are used for field generation: the single turn coil
technique \cite{Nakao, Miura} and the electromagnetic flux
compression method.\cite{Matsuda_EMFC, Miura} The single-turn coil
method can generate 200~T without any sample damage and thus we
can repeat measurements on the same sample under the same
experimental conditions. For higher field experiments we use the
electromagnetic flux compression method that can generate fields
up to 600~T. This is a totally destructive method and we lose the
sample as well as the magnet in each shot. The maximum energy
stored in capacitor banks is 200~kJ for the single-turn coil
method and 5~MJ for the electromagnetic flux compression method.

CR experiments were carried out by measuring the transmission of
midinfrared radiation through the sample as a function of magnetic
field.  We use several types of infrared lasers as radiation sources.
The wavelength range of the radiation was 28 $\mu$m to 3.39 $\mu$m,
corresponding to a photon energy range of 44~meV to 366~meV. The use of
the high magnetic fields required these unusually high photon energies
for CR. In addition, the photon energy had to be higher
than the plasma frequency; the samples were heavily doped and the plasma
frequency was estimated to be $\sim$30 meV and $\sim$60 meV for the light
holes and heavy holes using effective masses of 0.026$m_0$
and 0.35$m_0$, respectively, and a concentration of 1 $\times$ 10$^{19}$
cm$^{-3}$.

We measured six $p$-InMnAs samples with various Mn concentrations ($x$ =
0, 0.003, 0.0045, 0.006, 0.02, and 0.025). The samples were grown by low
temperature molecular beam epitaxy.  The characteristics of the
samples are shown in Table~\ref{tab:table1}. The hole concentration $n_p$,
the mobility $\mu$ and the resistivity $\rho$
shown were obtained by DC transport measurements at room temperature.
The dimensions of the sample for the megagauss CR experiment were about
2 $\times$ 2 mm$^2$ with $\sim$ 1 $\mu$m thickness.  The thickness of the
substrate (semi-insulating GaAs) was 0.5 mm. The temperature was controlled from
room temperature to 10 K using a liquid-helium flow cryostat made
of plastic.
\begin{table}
\caption{\label{tab:table1}Characteristics of the In$_{1-x}$Mn$_x$As samples used in the
present work. All samples are 1 $\mu$m thick. The $x$=0 sample (InAs) is Be doped.} 
\begin{ruledtabular}
\begin{tabular}{cccccc}
$x$  & buffer & $n_p$ &
$\mu$ & $\rho$ & $T_c$(K)\\
 & & (10$^{19}$ cm$^3$) &
(cm$^2$/Vs) & (m$\Omega$ cm) \\
\hline
0& - & 0.5 & 200 &
& - \\
0.0015& - & 1.1 & 59 &6.6
& - \\
0.003& - & 2.30 & 56 &4.9
& $< 4$ \\
0.0045& - & 2.3 & 56 &4.8
& $< 4$ \\
0.006&- & 1.8 & 64 &5.3
& $< 4$ \\
0.02 & - & 0.39 & 26 &62
& $\sim$ 4 \\
0.025 & InAs\footnotemark[1] & $\sim$ 1 &  &
& $< 10$ \\
\end{tabular}
\end{ruledtabular}
\footnotetext[1]{Thickness is 190 nm.}
\end{table}


\section{Experimental Results and Theoretical Analyses}
\label{sec:level3}
\subsection{Mn Concentration Dependence}
%
\begin{figure}
\begin{center}
\includegraphics[scale=0.55]{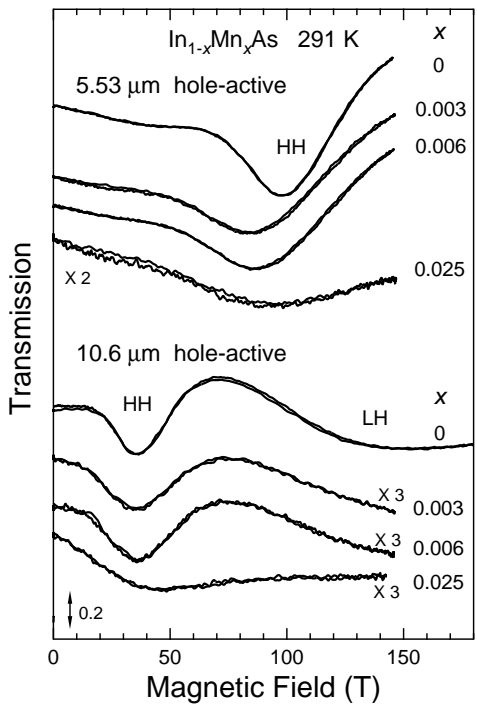}
\caption{\label{fig:CR_x} Room temperature cyclotron resonance spectra for
In$_{1-x}$Mn$_x$As with various $x$ taken with hole-active
circularly-polarized 5.53 $\mu$m and 10.6 $\mu$m radiation.}
\end{center}
\end{figure}
%
\begin{figure}[t]
\begin{center}
\includegraphics[scale=0.48]{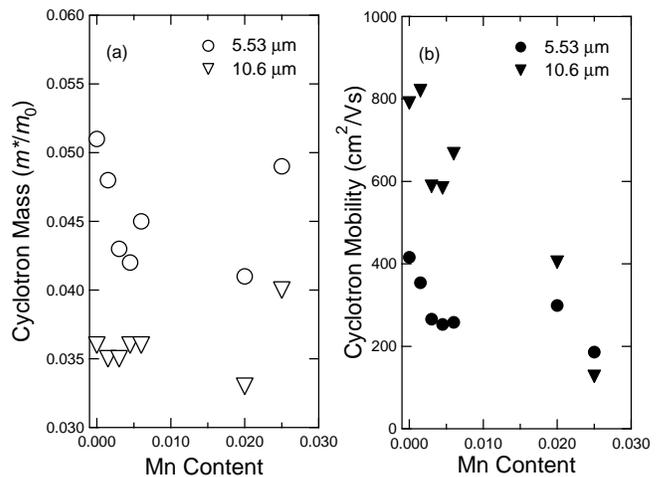}
\caption{Deduced cyclotron masses (a) and cyclotron mobilities (b)
at 291 K as functions of Mn content, $x$. Open circles and open triangles shows
cyclotron masses at 5.53 $\mu$m and 10.6 $\mu$m, respectively.
Closed circles and closed triangles denote the cyclotron mobilities at
5.53 $\mu$m and 10.6 $\mu$m, respectively.}
\label{summary}
\end{center}
\end{figure}
Typical room-temperature magneto-transmission spectra at 5.53 $\mu$m
($\hbar\omega$ = 0.224 eV) and 10.6 $\mu$m ($\hbar\omega$ = 0.117
eV) for different Mn concentrations are shown in Fig.~\ref{fig:CR_x}.
Quarter-wave plates were used to create a hole-active circularly polarized
beam.  For all samples, a broad absorption peak was observed at 80-100 T
for 5.53 $\mu$m and at $\sim$ 40 T for 10.6 $\mu$m.  Based on our detailed
theoretical analysis (to be discussed later), we attribute this peak to
heavy hole (HH) CR.  We observe another absorption band at $\sim$ 150 T
with 10.6 $\mu$m radiation, and attribute it to light hole (LH) CR.  The
cyclotron masses and cyclotron mobilites of HH were deduced from the peak
positions and widths, respectively, and plotted in Fig.~\ref{summary}.
Only the transition between Landau levels with small indices (quantum
numbers) contributes to the CR even at room temperature due to the
large cyclotron energy at high fields in the megagauss range.  Thus,
the obtained masses are significantly smaller than the band edge HH mass
(0.35$m_0$) due to the well known quantum effect in high magnetic fields.

As shown in Fig.~\ref{summary}(a), the HH cyclotron mass tends to decrease
with $x$ up to $x$ = 0.02 when measured at 5.53 $\mu$m.  However, the HH
mass for the $x$ = 0.025 sample apparently deviates from this tendency.
In addition, the HH mass obtained at 10.6 $\mu$m does not show any obvious
dependence on the Mn concentration.  It may suggest that some other
effects such as strain and disorder are playing a role in determining
the masses. Actually, as shown later, our detailed Landau level calculation 
at the high fields suggests that the HH band edge mass in In$_{1-x}$Mn$_x$As is 
nearly the same with the HH band edge mass in InAs.

The cyclotron mobilities in Fig.~\ref{summary}(b) are larger than
those obtained by DC transport measurements even in the nonmagnetic
($x$ = 0) sample.  This is partly because the HH cyclotron masses
in this high field range are an order of magnitude smaller than the
classical value, as we pointed out above.  Another possible origin  of
the discrepancy is the reduction of magnetic scattering at high fields,
i.e., the randomness of Mn spins decreases at strong magnetic
fields even at room temperature. This is consistent with the fact that
the relative increase of the mobility compared to the DC value is
larger for the magnetic samples than the nonmagnetic sample (InAs).
However, the fact that cyclotron mobilities measured at 10.6 $\mu$m
(resonance fields $\sim$40 T) are larger than those measured at 5.53
$\mu$m (resonance field $\sim$ 80-100 T) is not explainable by this
picture.  Since at room temperature the magnetization is not saturated even at 100 T, 
the randomness of Mn spins should be smaller for
higher fields.\cite{MatsuPhysB} Although we take into account the mass
increase due to the band non-parabolicity, the scattering time ($\tau$)
deduced from the cyclotron mass ($m_{CR}$) and the cyclotron mobility
($\mu^*=e\tau/m_{CR}$) is found to be smaller at 5.53 $\mu$m than that
at 10.6 $\mu$m.  Moreover, the cyclotron mobility is not dependent on
the temperature as we show later.  Hence, it is expected that magnetic
effects on scattering processes are suppressed at lower fields than the
field range where we observed CR.
\subsection{Temperature Dependence}
%
%
\begin{figure}
\begin{center}
\includegraphics[scale=0.55]{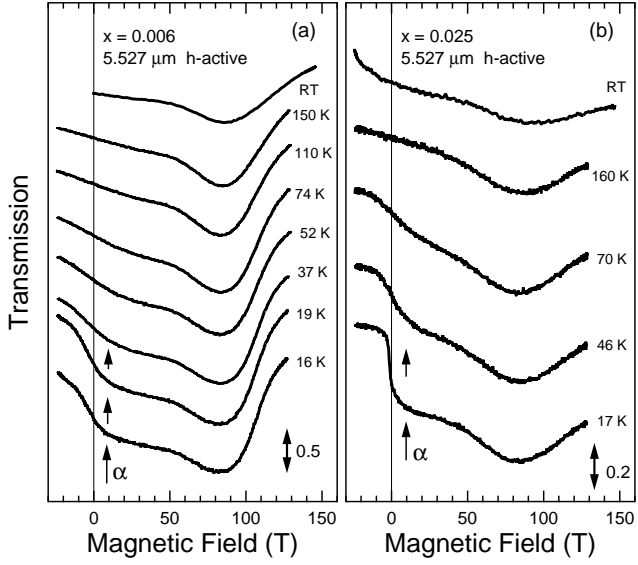}
\caption{Temperature-dependent
cyclotron resonance traces for In$_{1-x}$Mn$_x$As with (a) $x$ = 0.006
and (b) $x$ = 0.025.
A new absorption band, labeled $\alpha$, appears at low temperatures.
}
\label{temp1}
\end{center}
\end{figure}
%
\begin{figure}
\begin{center}
\includegraphics[scale=0.6]{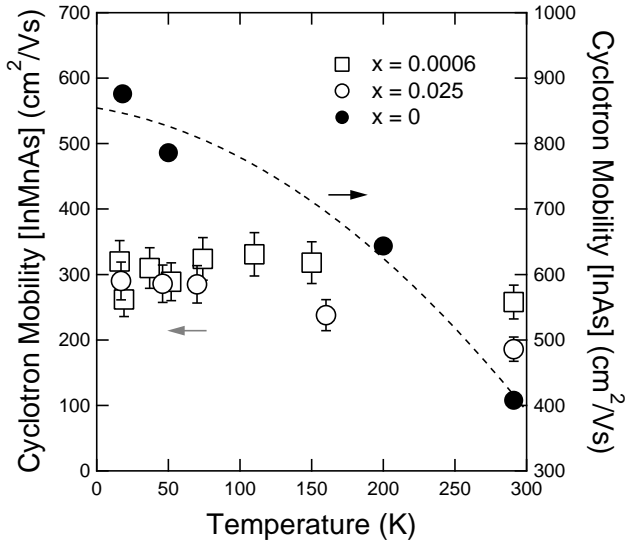}
\caption{Temperature dependence of cyclotron mobilities of
heavy hole CR at 10.6 $\mu$m for In$_{1-x}$Mn$_x$As ($x$ = 0, 0.006, 0.025).
Dashed curve is a guide for the eyes for the InAs data.
}
\label{temp2}
\end{center}
\end{figure}

CR spectra for the $x$ = 0.006 and $x$ = 0.025 samples at various
temperatures are shown in Figs.~\ref{temp1}(a) and \ref{temp1}(b),
respectively.  The HH CR peak position is found to be rather insensitive
to the temperature in both samples, suggesting that any effect of
tempereture-dependent exchange splitting of the Landau levels is
small. This is because the Mn concentration is small and the relative
energy of the exchange splitting to the cyclotron energy is small in
the megagauss range.  Also, we find that the cyclotron mobility shows
no significant change for both samples.  This should be contrasted to
the cyclotron mobility of InAs, which increases with decreasing
temperature, as shown in Fig.~\ref{temp2}.  It is likely that impurity
scattering is more significant in InMnAs than InAs and the effect of
suppression of magnetic scattering at low temperature which is expected
for InMnAs, is not large enough to be observed.

\begin{figure}[b]
\includegraphics[scale=0.57]{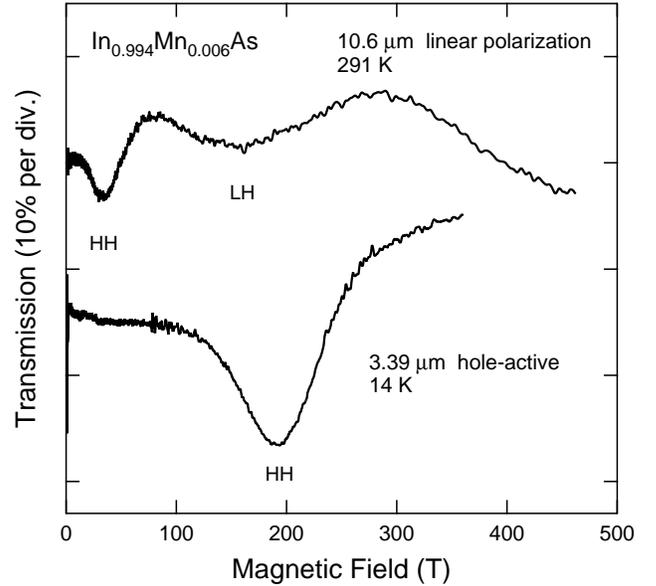}
\caption{Cyclotron resonance in In$_{1-x}$Mn$_x$As ($x$ = 0.006) at multi-megagauss
fields. Only the HH CR is observed at 3.39 $\mu$m and 14 K. The absorption band 
observed at low fields and low temperatures for the wavelength of 5.53 $\mu$m (the band labeled $\alpha$) is not observed.
}
\label{MegaCR}
\end{figure}

It should be noted that a new broad absorption band appears at low temperature
and low fields below 20 T for both samples, Labeled `$\alpha$' in
Fig.~\ref{temp1}.  One of the possible origins of this band is impurity
shifted CR (ICR) absorption. If this is the case, it could be closely
related to the Mn impurity band and useful for clarifying the picture of
ferromagnetism.  However, our CR experiments at higher photon energies
(see, e.g., the trace for 3.39 $\mu$m and 14 K in Fig.~\ref{MegaCR})
and theoretical calculations indicate that this is not the case.

As we decribed previously,\cite{Sanders,Giti}
the width of the low field tail results from higher order Landau
levels being populated and thus it depends on
the free hole concentration.  In addition, the sharpness
of the low field cutoff is seen to depend on temperature and
can be attributed to the sharpness of the Fermi distribution at
low temperatures. The results show that the cyclotron resonance
for the InMnAs case is made up of one strong heavy
hole transition together with background absorption due to
higher populated Landau levels at low values of the magnetic
field. Hence, it is suggested that the electronic structure in In$_{1-x}$Mn$_x$As 
is not strongly modified from that in InAs. Actually, our Landau level calculation 
using the Luttinger parameters for InAs can reproduce very well the CR absorption curves as shown 
later. (See Fig.~\ref{sim_CR}.)
%
\subsection{Photon Energy Dependence}
\begin{figure}
\includegraphics[scale=0.5]{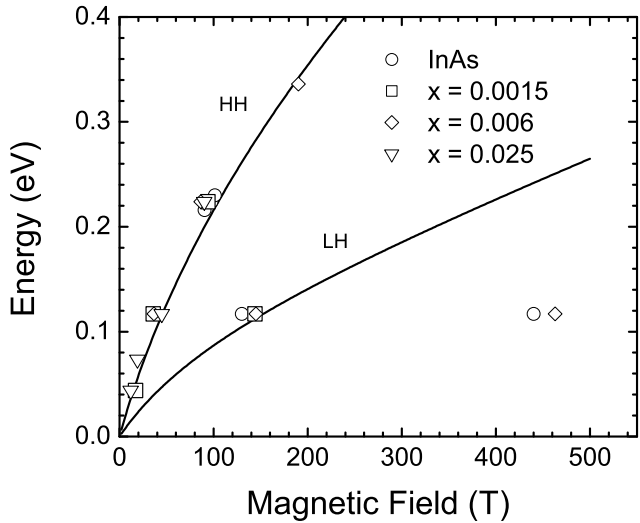}
\caption{Summary of the observed hole-active cyclotron resonance peak positions
for four samples with different Mn contents.
The solid curves are calculations based on an 8-band effective mass model.
for bulk InAs.}
\label{En_Field}
\end{figure}

Figure~\ref{En_Field} summarizes the observed hole-active CR peak
positions as a function of magnetic field (B).
The solid curves show the calculated inter-Landau-level
transitions for bulk InAs based on an 8-band Pidgeon-Brown
model.\cite{nIMA-CR,Sanders,Sanders2,Sanders3,Sun1}
The details of the calculations in terms of the
effective mass theory is described in the next subsection. Although the
effective mass theory could break down at such high fields since the
magnetic length becomes $\sim$ 12 {\AA} at 450 T, which is not too large
compared to the lattice constant, we see that most of the CR transitions
are explained well.
%
\subsection{Landau Level Calculation}

We have performed detailed calculations of the valence-band
Landau levels within an 8-band {\bf k$\cdot$p}, Pidgeon-Brown
model\cite{PB,Sanders2,Sanders3} including {\it sp-d} coupling to the Mn
ions to understand the observed CR.  The Landau levels and the possible
inter-Landau-level transitions are complicated at high fields due to
wave function mixing.
%
%
\begin{figure}[b]
\includegraphics[scale=0.80]{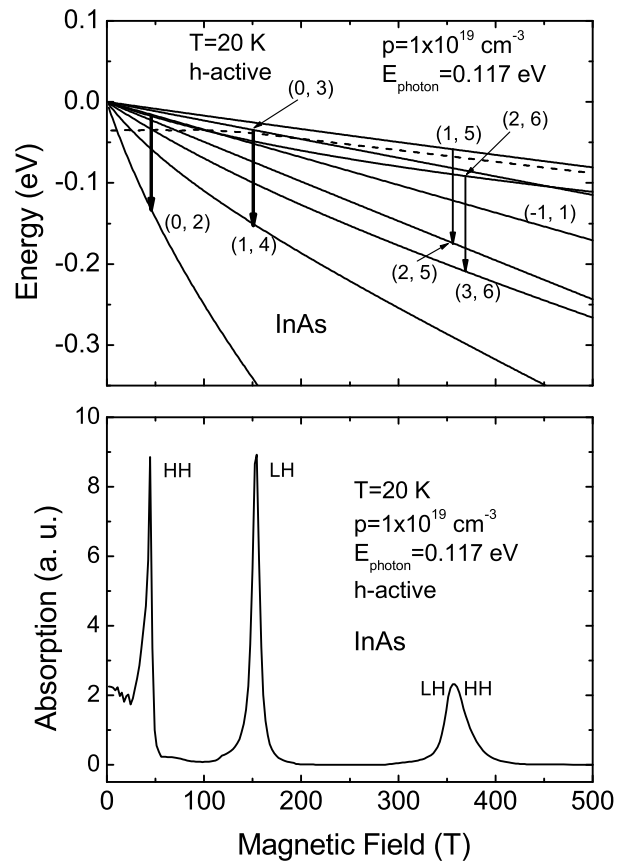}
\caption{Calculated Landau levels of the valence band at $\Gamma$
point and CR absorption as functions of magnetic field for $x$ = 0
at 20 K. The Fermi energy corresponding to the hole concentration
1 $\times$ 10$^{19}$ cm$^{-3}$ is also shown as the dashed line.
The Lower panel shows CR spectrum which corresponds to the inter-Landau
level transitions shown in upper panel.}
\label{LL_cal}
\end{figure}
%
%
%
\begin{figure}[b]
\includegraphics[scale=0.5]{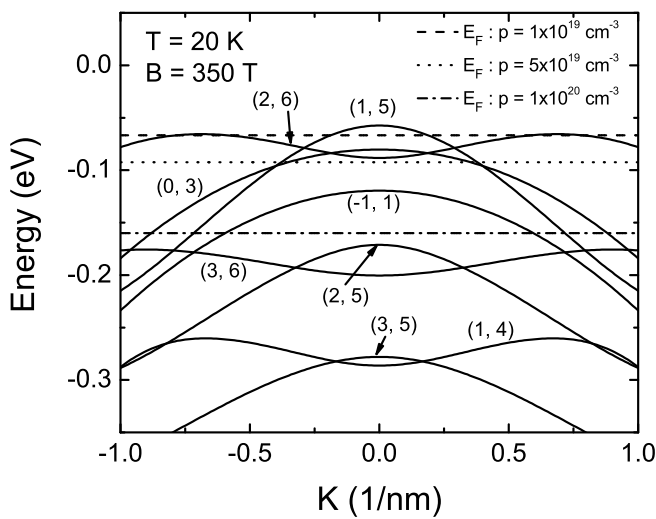}
\caption{Calculated energy $E$ versus wavevector $k$ (in the magnetic
field direction) for B = 350 T. As can be seen, the maximum in the (2,6)
Landau Level does not occur at $k=0$.  For higher carrier densities, the
appropriate Fermi levels are shown.  Higher carrier densities can lead
to a much broader absorption feature.}
\label{klevel2}
\end{figure}

Figure \ref{LL_cal} shows the calculated Landau levels at the $\Gamma$
point ({\bf k} = 0) as a function of magnetic field for $x$ = 0 at 20 K.
The Fermi energy corresponding to a hole concentration of $n_p = 1 \times
10 ^{19}$ cm$^{-3}$ is also indicated in the figure (dashed line). In the
figure, we use the notation ($n$,$m$) to denote the levels.  $n$ corresponds to
the manifold index ($n$ = $-$1,0,1,\dots) and $m$ denotes the state within the
manifold (up to 8 states for manifolds $n$ = 2 and higher).  This notation 
is basically the same with the notation given by Pidgen and Brown \cite{PB} and 
discussed in detail in the article by Kossut for narrow gap II-VI diluted magnetic 
semiconductors.\cite{Kos88183}

We used standard Luttinger parameters for InAs and took into account
the $sp-d$ exchange interactions for calculations involving Mn doping.
In this case, the exchange constants we used for the electron-Mn
interaction ($\alpha$) and hole-Mn ion ($\beta$) were 0.5 eV and $-$1.0
eV, respectively.\cite{nIMA-CR}

Allowed optical transitions for h-active polarization and a laser energy of
0.117 eV, or wavelength of 10.6 $\mu$m (top curve in Fig.~\ref{MegaCR})
are shown.  Analysis of the wave functions for the first transition
($\approx$ 50 T) show that this corresponds to a transition between the 
lowest and the second lowest spin down HH Landau levels (i.e., the dominant components
of the wavefunctions are HH spin down).  The next transition
($\approx$ 150 T) occurs between the lowest and the second lowest spin down LH Landau
levels. We thus label the experimental dips in the transmission (peaks
in the absorption) HH and LH, respectively, in Figs.~\ref{MegaCR} and
~\ref{LL_cal}.  We note
that for low fields below 50 T, Fig.~\ref{LL_cal} shows absorption,
though not any peak structure.  This corresponds to transitions from
the many higher order
Landau levels which lie below the Fermi energy at low magnetic fields.
A similar feature is seen in the experimental data of Fig.~\ref{MegaCR}.

At very high fields ($>$ 300 T) the experimental data shows additional CR
absorption (c.f. Fig.~\ref{MegaCR}).  This is also seen in the calculated
CR absorption in Fig.~\ref{LL_cal}.  Our calculations show that this
feature arises from multiple transitions, one being LH spin up like
and the other being HH spin up like.  While it appears that the (2,6) HH
like transition lies below the Fermi level, the dominant part of this
transition does not occur at $k=0$.  Figure \ref{klevel2} shows the $E$
vs. $k$ plot for T = 20 K and B = 350 T.   As can be seen, the maximum
of the (2,6) band occurs away from $k=0$, and the $k = 0 $ point is not
occupied at a density of $p$ = 1 $\times$ 10$^{19}$ cm$^{-3}$.  Our calculations
show that the optical matrix element for the (2,6) $\rightarrow$ (3,6)
transition is strongest at the maximum point away from $k=0$ and that
for higher carrier densities, the CR absorption broadens to higher fields.

\subsection{Electron-Active Hole Cyclotron Resonance}
Hole CR can be observed not only with hole-active circular polarized
radiation but also with electron-active circularly polarization due to the
complex Landau level formation.\cite{Sun1}
Reversing the order of the Landau level index in the energy diagram for some
Landau levels allows $N+1 \rightarrow N$
type absorption transition that is active for the opposite circular
polarization (electron active
circular polarization) to the hole CR transition ($N \rightarrow N+1$), where $N$
denotes the Landau level index.

\begin{figure}[t]
\includegraphics[width=7cm]{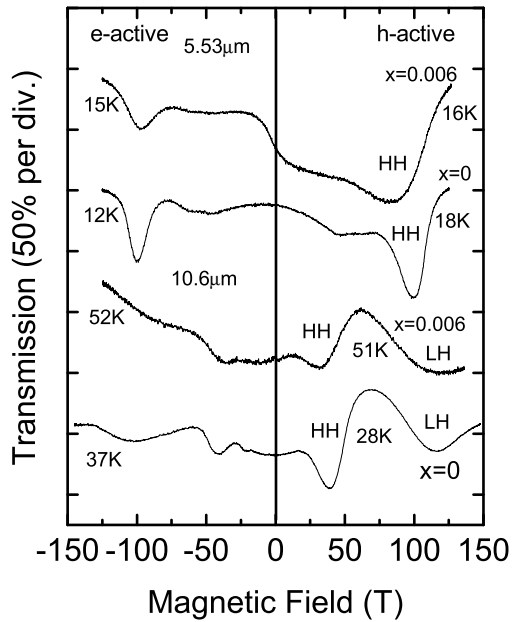}
\caption{Cyclotron resonance in  In$_{1-x}$Mn$_x$As ($x$=0, 0.006) at 10.6 $\mu$m (upper part of the
figure) and at 5.53 $\mu$m
(lower part of the figure). The right hand side of the figure corresponds to the results using the hole
active circularly polarization of the radiation and the left hand side of the figure shows the results
using the electron active circularly polarization. }
\label{eh_CR}
\end{figure}

In Fig.~\ref{eh_CR} electron active spectra are represented by reversing
the sign of the magnetic field.
Although the spectra look complicated, we can obtain rich information on
the valence band parameters from the spectra. To understand details of the CR
spectra, we have
simulated the CR spectra using the Fermi's golden rule and the calculated
Landau levels.

\begin{figure}
\includegraphics[width=7cm]{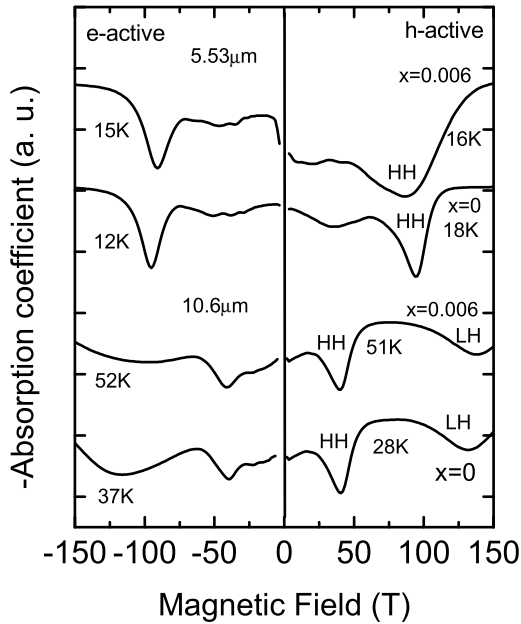}
\caption{Calculated CR spectra which can be directly compared
to the experimental results shown in
Fig.~\ref{eh_CR}.}
\label{sim_CR}
\end{figure}

Figure~\ref{sim_CR} shows the result of the simulation and can be directly
compared to Fig.~\ref{eh_CR}.
We see that the theory
not only explains the CR peak positions but also the relative absorption
intensity very well for both samples
($x$ = 0 and $x$ = 0.006). This fact suggests that the theoretical treatment,
taken in this work, i.e., the effective
mass theory and mean field approximation, is adequate for describing the
electronic structure of the
InAs-based DMS and the difference of the electronic structure between InAs
and InMnAs is not very large when
the Mn concentration is as small as 1\%.
%
%
%

%
%
%
\section{conclusions}
\label{sec:level4}

Ultrahigh magnetic field cyclotron resonance experiments in $p$-type
InMnAs thin films have revealed that not all of the holes are localized.
The cyclotron resonance experiments are consistent with the delocalized holes
with the effective masses given by the band structure (0.35$m_0$ for the heavy holes 
at the band edge).  
Our detailed
theoretical treatment using the effective mass theory within a mean
field approximation for the polarization of the Mn ions successfully
reproduced the complicated CR spectra. This fact strongly suggests that
the mean-field approximation is adequate for the narrow gap InMnAs DMS
and that the ferromagnetic transition can be understood in terms of
Zener's $p-d$ exchange interaction rather than the double exchange interaction. Although the present experiments are
only for $T > T_c$,  recent experiments in InMnAs/GaSb heterostructure
samples below the ferromagnetic transition temperature\cite{Giti} have
confirmed the existence of delocalized holes in the ferromagnetic InMnAs
DMS system.

We note, however, that hole CR has {\it not} been observed in ferromagnetic
GaMnAs even in the megagauss range.\cite{MatsuPhysB,pasps3} Moreover, it
has been reported that the optical conductivity spectrum shows non-Drude
type behavior in GaMnAs\cite{Singley, Hirakawa2, Nagai} while it
clearly shows Drude type conductivity in InMnAs.\cite {Hirakawa1}. These
facts suggest that the hole character may be different in InMnAs and
GaMnAs. According to the recent report on ZnCrTe,\cite{Saito,Saito2}
ferromagnetism can take place even if the conductive hole
concentration is extremely low ($\sim$ 10$^{15}$ cm$^{-3}$).  Therefore, it
is likely that the ferromagnetic transition mechanism is material dependent
and strongly depends on the electronic states at the Fermi energy. We can
imagine that the $d$-holes participate in the ferromagnetic
transition mechanism in ZnCrTe and the nature of the hole states in GaMnAs 
falls between InMnAs and ZnCrTe.

We have shown that ultrahigh-field CR experiments are
one of the most powerful methods to clarify the nature of carries in the
ferromagnetic semiconductors. It would be intriguing to
perform CR experiments in a material with a moderate degree
of localization between InMnAs and GaMnAs.
%
%
%
\section{Acknowledgements}
\label{sec:level6}
This work was supporetd by the NEDO International Joint Research
Program and DARPA MDA972-00-1-0034 (SPINS), NSF DMR-0134058,
NSF DMR-9817828, NSF INT-0221704.
%
%

 \end{document}